\documentclass[reprint]{revtex4-1}

\usepackage{array}
\usepackage{graphicx}
\usepackage{verbatim}
\usepackage{color}
\usepackage{multirow}
\usepackage{longtable}

\newcommand{\bra}[1]{\left\langle{#1}\right\vert}
\newcommand{\ket}[1]{\left\vert{#1}\right\rangle}

\begin{document}

\title{Multilevel distillation of magic states for quantum computing}
\author{Cody Jones}
\email{ncodyjones@gmail.com}
\affiliation{Edward L. Ginzton Laboratory,
         Stanford University,
         Stanford, California 94305-4088, USA}

\begin{abstract}
We develop a procedure for distilling magic states used in universal quantum computing that requires substantially fewer initial resources than prior schemes.  Our distillation circuit is based on a family of concatenated quantum codes that possess a transversal Hadamard operation, enabling each of these codes to distill the eigenstate of the Hadamard operator.  A crucial result of this design is that low-fidelity magic states can be consumed to purify other high-fidelity magic states to even higher fidelity, which we call ``multilevel distillation.''  When distilling in the asymptotic regime of infidelity $\epsilon \rightarrow 0$ for each input magic state, the number of input magic states consumed on average to yield an output state with infidelity $O(\epsilon^{2^r})$ approaches $2^r+1$, which comes close to saturating the conjectured bound in [Phys. Rev. A 86, 052329].  We show numerically that there exist multilevel protocols such that the average number of magic states consumed to distill from error rate $\epsilon_{\mathrm{in}} = 0.01$ to $\epsilon_{\mathrm{out}}$ in the range $10^{-5}$ to $10^{-40}$ is about $14\log_{10}(1/\epsilon_{\mathrm{out}}) - 40$; the efficiency of multilevel distillation dominates all other reported protocols when distilling Hadamard magic states from initial infidelity 0.01 to any final infidelity below $10^{-7}$.  These methods are an important advance for magic-state distillation circuits in high-performance quantum computing, and they provide insight into the limitations of nearly resource-optimal quantum error correction.
\end{abstract}

\maketitle

\section{Introduction}
Quantum computing can potentially solve a handful of otherwise intractable problems, such as factoring large integers~\cite{Shor1999} or simulating quantum physics~\cite{Lloyd1996}.  Though the number of applications with a known ``quantum speed-up'' is small, some are quite valuable, like the preceding examples.  Quantum computations depend on coherent entangled states which are very sensitive to noise, so fault-tolerant quantum computing addresses imperfections in physical hardware with error-correcting codes~\cite{Preskill1998,Nielsen2000}, the most studied of which are stabilizer codes~\cite{Gottesman1997}.  However, while quantum codes protect against noise, no code natively supports a universal set of transversal gates for simulating any quantum circuit~\cite{Zeng2007,Eastin2009}.  To achieve universal quantum computing with error correction, Bravyi and Kitaev proposed a solution~\cite{Bravyi2005} that has received considerable attention: inject faulty ``magic states'' into the code, purify them using the error-corrected gates, then consume them to implement otherwise unavailable quantum circuits.  These states are ``magic'' because it is possible to distill a subset of high-fidelity states from an ensemble of faulty states and because they enable universal fault-tolerant quantum computation.

Magic state distillation has been the subject of intense investigation in recent years.  Knill independently introduced a distillation procedure for $\ket{H}$, the (+1) eigenstate of the Hadamard operation~\cite{Knill2004}, prior to the work by Bravyi and Kitaev~\cite{Bravyi2005}.  Reichardt showed that these protocols were equivalent and introduced an improvement which increased the threshold error rate~\cite{Reichardt2005}.  More recently, Meier~\emph{et al}. introduced a 10-to-2 distillation procedure based on a code with two encoded qubits~\cite{Meier2012}, and Bravyi and Haah introduced a ($3k+8$)-to-$k$ procedure using so-called triorthogonal codes with even $k$ encoded qubits~\cite{Bravyi2012}.  The distillation procedures we develop herein continue this trend of using larger, multi-qubit codes.  The relationship of magic-state distillation to other aspects of quantum information has also been an area of active study.  Fowler and Devitt have proposed methods to reduce the size of distillation circuits when using topological quantum error correction~\cite{Fowler2012}.  Magic-state distillation has been demonstrated experimentally in NMR~\cite{Souza2011}.  Additionally, distillation protocols for qudits have been proposed and analyzed~\cite{Campbell2012,Veitch2012}.

For a quantum state, we quantify the probability of it having an error using the infidelity $1-F$, where $F = \bra{\psi}\rho\ket{\psi}$ is the fidelity between some mixed state $\rho$ and the ideal state $\ket{\psi}$.  The initial $\ket{H}$ states are prepared in a faulty manner before being injected into a fault-tolerant quantum code, and Reichardt proved that the theoretical-limit infidelity for $\ket{H}$ states to be distillable is about $0.146$~\cite{Reichardt2005}.  Campbell and Browne examined further properties of mixed states that may be distilled~\cite{Campbell2010}.  The efficiency of distilling high-infidelity magic states to low infidelity is of great importance to fault-tolerant quantum computing.  Although magic states are the widely preferred method for achieving universality, distillation circuits are currently estimated to require the majority of resources in a quantum computer~\cite{Isailovic2008,Jones2012}.  Therefore, advances in distillation protocols are important steps toward making quantum computing possible.

This paper presents two important, related results.  First, we specify a family of $[[n,(n-4),2]]$ Calderbank-Shor-Steane (CSS) quantum stabilizer codes~\cite{Calderbank1996,Steane1996} known as ``$H$~codes'' with transversal Hadamard operation, for even $n \ge 6$.  A ``transversal'' quantum operation is one where a gate acting on a logical, encoded qubit is implemented by independent gates on each qubit in that code block (see p. 483 of Ref.~\cite{Nielsen2000}).  The $H$~codes are dense, because the ratio of logical qubits to physical qubits $(n-4)/n \rightarrow 1$ as $n \rightarrow \infty$.  Second, we demonstrate that concatenated versions of $H$~codes allow for distillation of high-fidelity encoded magic states by consuming low-fidelity magic-state ancillas.  We call this ``multilevel distillation,'' because each such protocol takes two types of magic-state inputs, which have different levels of infidelity and which are applied at different concatenation levels within the distillation circuit.  Multilevel protocols lead to the most efficient procedure for distilling magic states reported so far.  For suitably small infidelity $\epsilon$ in each input magic state with independent errors, there exists a sequence of multilevel protocols that yields output magic states with infidelity $O(\epsilon^{2^r})$ and requires asymptotically $2^r+1$ input states per distilled output state.  This efficiency comes close to the ``optimality'' bound conjectured in Ref.~\cite{Bravyi2012}.  For the purposes of developing quantum devices, this result is useful for exposing limits for optimizing quantum error correction.  While this result is interesting theoretically, we also numerically study the distillation efficiency for $\epsilon_{\mathrm{in}} = 0.01$, which is of practical importance to fault-tolerant quantum computing.  We find that multilevel distillation is superior to previously reported protocols when the final infidelity is below $10^{-7}$.

Throughout this paper, we adopt the following notation for single-qubit Pauli operators, for readability: $\texttt{X} \equiv \sigma^x$, $\texttt{Z} \equiv \sigma^z$, and $\texttt{I}$ is the identity operator on a qubit.  Additionally, we use ``physical qubit'' to denote those qubits used to produce a quantum code, whereas ``logical qubits'' are the protected information inside the code, again for readability.  It may be the case that physical qubits for one encoding level are themselves the logical qubits of another code, which is a standard technique of quantum code concatenation~\cite{Knill1996,Preskill1998,Nielsen2000}.

\section{A family of codes with transversal Hadamard}
We define a family of CSS quantum codes which encode an even number $k$ logical qubits using $(k+4)$ physical qubits and possess a transversal Hadamard operation, so we call them collectively ``$H$~codes'' and denote $H_n$ as the code using $n = k+4$ physical qubits.  Any $H$~code may be defined as follows.  The stabilizer generators are $S_1 = \texttt{X}_1 \texttt{X}_2 \texttt{X}_3 \texttt{X}_4$, $S_2 = \texttt{Z}_1 \texttt{Z}_2 \texttt{Z}_3 \texttt{Z}_4$, $S_3 = \texttt{X}_1 \texttt{X}_2 \texttt{X}_5 \texttt{X}_6 \ldots \texttt{X}_n$, $S_4 = \texttt{Z}_1 \texttt{Z}_2 \texttt{Z}_5 \texttt{Z}_6 \ldots \texttt{Z}_n$, where subscripts index over physical qubits and tensor product between Pauli operators is implicit.  The logical Pauli operators (corresponding to logical qubits), denoted with an over bar and indexed by $i = 1 \ldots k$, are $\overline{\texttt{X}}_i = \texttt{X}_1 \texttt{X}_3 \texttt{X}_{i+4}$ and $\overline{\texttt{Z}}_i = \texttt{Z}_1 \texttt{Z}_3 \texttt{Z}_{i+4}$.  The Hadamard transform exchanges \texttt{X} and \texttt{Z} operators, so application of transversal Hadamard gates at the physical level enacts a transversal Hadamard operation at the logical level, which will be a useful property when we later concatenate these codes.  All $H$~codes have distance two, which means they can detect a single physical Pauli error.  The product of two logical Pauli operators of the same type for two distinct logical qubits has weight two (number of non-identity physical, single-qubit Pauli operators); the product of same-type Pauli operators on all logical qubits is also weight-two at the physical level.  The stabilizers come in matched \texttt{X}/\texttt{Z} pairs, so there are no weight-one logical operators.

\begin{figure}
  \centering
  \includegraphics[width=8.3cm]{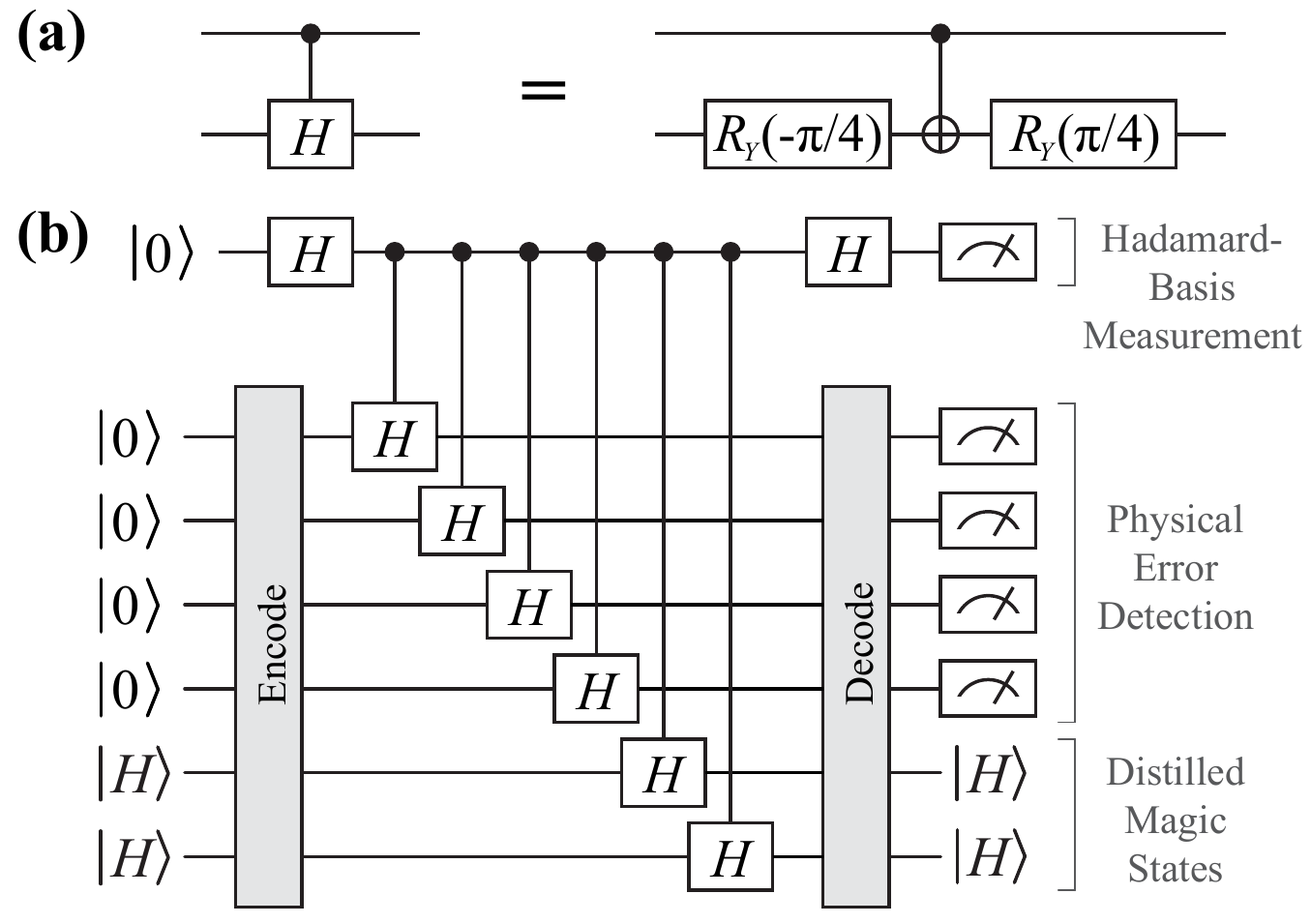}\\
  \caption{Distillation of $\ket{H}$ magic states using an $H$~code.  \textbf{(a)}~Controlled-Hadamard gate is constructed using $R_Y(-i\pi/4) = \exp(i\pi\sigma^Y/8)$ and its inverse, each of which requires one $\ket{H}$ state~\cite{Meier2012}.  \textbf{(b)}~Initial $\ket{H}$ states (left) are encoded with four additional qubits, initialized to $\ket{0}$ here.  The boxes ``Encode'' and ``Decode'' represent quantum circuits for encoding and decoding, which are not shown here.}
  \label{Had_distill_figure}
\end{figure}

The $(+1)$~eigenstate $\ket{H} = \cos(\pi/8)\ket{0} + \sin(\pi/8)\ket{1}$ of the Hadamard operator $\texttt{H} = (1/\sqrt{2})(\texttt{X} + \texttt{Z})$ is a magic state for universal quantum computing~\cite{Knill2004,Bravyi2005,Reichardt2005,Meier2012,Bravyi2012}.  In particular, two of these magic states can be consumed to implement a controlled-$\texttt{H}$ operation~\cite{Knill2004,Meier2012}, enabling one to measure in the basis of $\texttt{H}$ (see Fig.~\ref{Had_distill_figure}a).  Our distillation procedure is as follows: (a)~encode faulty $\ket{H}$ magic states in an $H$~code; (b)~measure in the basis of the transversal Hadamard gate by consuming $\ket{H}$ ancillas; (c)~reject the output states if either the measure-Hadamard or code-stabilizer circuits detect an error.  For example, when an $H_{(k+4)}$~code is used for distillation, $k$ $\ket{H}$ states are encoded as logical qubits using $(k+4)$ physical qubits.  Each transversal controlled-Hadamard gate consumes two $\ket{H}$ states~\cite{Meier2012}, and this gate is applied to all physical qubits, which results in the $(3k+8)$-to-$k$ input/output distillation efficiency of these codes.  A diagram of the quantum circuit for distillation using $H_6$ is shown in Fig.~\ref{Had_distill_figure}b.

\section{Multilevel distillation}
Multilevel distillation uses concatenated codes with transversal Hadamard for distillation, in such a manner that the protocol takes as input magic states at two different levels of infidelity, and the two types of magic states enter at different concatenation levels in the code.  The $\ket{H}$ ancillas consumed for transveral controlled-Hadamard measurement are of lower fidelity than the encoded logical $\ket{H}$ states being distilled.  When two quantum codes with transversal Hadamard are concatenated, the resulting code also has transversal Hadamard.  Under appropriate conditions, the distance of the concatenated code is the product of the distances for the individual codes: $d' = d_1 d_2$~\cite{Meier2012}.  Thus the concatenation of two $H$~codes yields a distance-4 code with transversal Hadamard, and $r$-level concatenation has distance $2^r$.

\begin{figure}
  \centering
  \includegraphics[width=8.3cm]{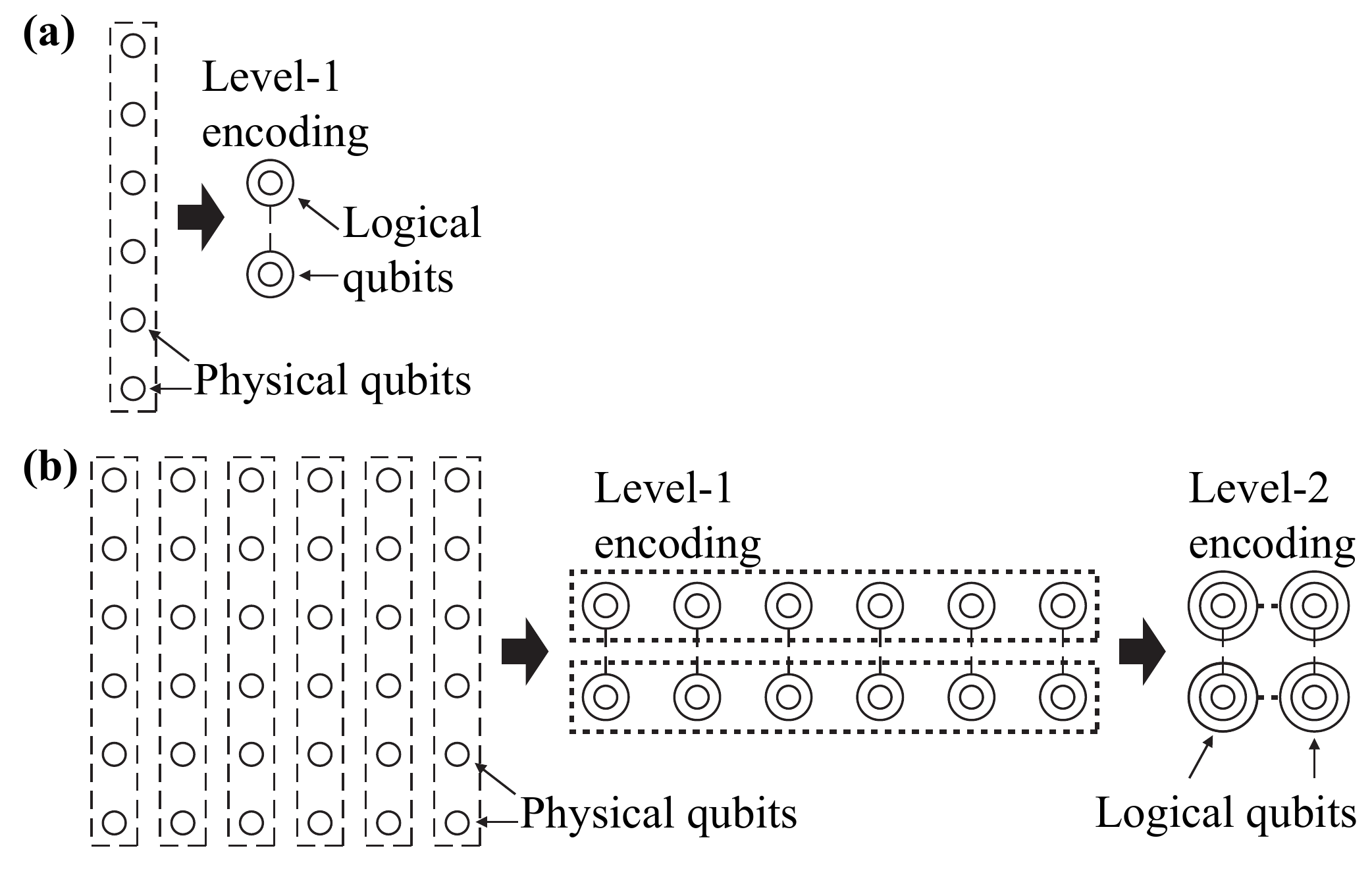}\\
  \caption{Concatenation of $H$~codes.  \textbf{(a)} Six physical qubits are coupled into an $H_6$~code with two logical qubits  \textbf{(b)} A $6 \times 6$ array of physical qubits are coupled into a concatenated two-level $H_6$~code.}
  \label{H_code_concatenation}
\end{figure}

The concatenation conditions for $H$~codes are that, through all levels of concatenation, any pair of physical qubits have at most one encoding block (at any level) in common.  The reasons for this restriction are that logical errors in the same block are correlated and that the statement above that distance multiplies through concatenation assumes independence of errors, so two qubits from the same encoding block can never be paired again in a different encoding block.  The required arrangement of qubits can be given a geometric interpretation.  Arrange all physical qubits at points on a cartesian grid in the shape of a rectangular solid, with the number of dimensions given by the number of levels of concatenation.  A square, cube, or hypercube are possible examples at dimensionality two, three, or four.  Each dimension is associated with a level of concatenation, and there must be an even $n \ge 6$ qubits in each dimension to form an $H$~code.  Construct $H$~codes in the first dimension by forming an encoding block with each line of qubits in this direction, as in Fig.~\ref{H_code_concatenation}a.  This will give rise to $k = n-4$ logical qubits along each line in this direction.  Repeat this procedure by grouping these first-level logical qubits in lines along the second dimension to form logical qubits in a two-level concatenated code, as in Fig.~\ref{H_code_concatenation}b.  Continuing in this fashion through all dimensions ensures that any pair of qubits have at most one encoding block in common.

As with the $H$~codes, multilevel codes use a transversal logical Hadamard-basis measurement to detect whether any one encoded qubit has an error (an even number of encoded errors would not be detected).  If the logical $\ket{H}$ states have independent error probabilities $\epsilon_l$, then the distilled states will have infidelity $O({\epsilon_l}^2)$ with perfect distillation.  We must also consider whether the Hadamard-basis measurement has an error.  For a two-level code arranged as a square of side length $n$, the transversal controlled-Hadamard gates at the lowest physical level require $(2 n^2)$ $\ket{H}$ magic states, each of which has infidelity $\epsilon_p$.  However, this is a distance-4 code, so for independent input error rates, the probability of failing to detect errors at the physical level is $O({\epsilon_p}^4) + O(\epsilon_l{\epsilon_p}^2)$ (rigorous analysis is provided later).  The code can detect more errors in the magic states at the lower physical level, so these $\ket{H}$ states can be of lower fidelity than the magic states encoded as logical qubits and successfully perform distillation.  This is the essential distinction between multilevel distillation and all prior distillation protocols.  When multiple rounds of distillation are required~\cite{Jones2012}, low-fidelity magic states are less expensive to produce, so multilevel protocols achieve higher efficiency.

Multilevel distillation protocols are applied in rounds, beginning with a small protocol (such as an $H$~code) and progressing to concatenated multilevel codes.  Let us denote the output infidelity from a single round by the function $\epsilon_{\mathrm{out}} = E_t^{n_1 \times \ldots n_t}(\epsilon_l,\epsilon_p)$.  For each such function, $t$ is the dimensionality (number of levels of concatenation) and $n_1 \ldots n_t$ are the sizes of each dimension, which need not all be the same.  As before, $\epsilon_l$ and $\epsilon_p$ refer to the independent error probabilities on logical and physical magic states, respectively.  A typical progression of rounds using a source of $\ket{H}$ states with infidelity $\epsilon_0$ might be $\epsilon_1 = E_1^n(\epsilon_0,\epsilon_0)$, $\epsilon_2 = E_2^{n \times n}(\epsilon_1,\epsilon_0)$, \emph{etc}.

Multilevel distillation circuits tend to be much larger in both qubits and gates than other protocols.  Because there can be many encoded qubits, the protocol is still very efficient, but the size of the overall circuit may be a concern for some quantum computing architectures.  At any number of levels, the distilled output states have correlated errors, so distilled magic-state qubits in our protocol must never meet again in a subsequent distillation circuit (we require that errors are independent within the same encoding block, as in Refs.~\cite{Meier2012,Bravyi2012}).  Let us suppose that one performs two rounds of distillation, where the first round uses one-level distillers with $k$ encoded magic states and the second round uses two-level distillers with $k^2$ encoded states.  Because the inputs to each distiller in the second round must have independent errors, there must be $k^2$ independent distillation blocks in the first round.  Therefore, to distill $k^3$ output states through two rounds, we require: $k^3 [\textrm{logical inputs}] + 2k^2(k+4) [\textrm{physical inputs}] + 2k(k+4)^2 [\textrm{physical inputs}] = 5k^3 + 24k^2 + 32k$ input states.

Consider a similar sequence through $r$ rounds with each distiller in round $q$ having $k^q$ encoded qubits.  The total number of \emph{logical} magic states is $k^r \times k^{r-1} \times \ldots k = k^{r(r+1)/2}$ to ensure that errors are independent between logical magic states in every round.  In the first round, the number of consumed magic states is $2(k+4)k^{r(r+1)/2 - 1}$; in any subsequent round $q \ge 2$, the number of consumed magic states is $2^{q-1} (k+4)^q k^{r(r+1)/2 - q}$ (recall that the Hadamard measurement is implemented $2^{q-2}$ times, meaning it is repeated for $q \ge 3$).  The total number of input magic states can thus be expressed as
\begin{equation}
\left[1 + \frac{k+4}{k} + \sum_{q=1}^r 2^{q-1}\left(\frac{k+4}{k}\right)^q\right] k^{r(r+1)/2}.
\end{equation}
For $r = 2$, this reproduces the expression above.  What also becomes clear is that the total size of multilevel protocols becomes unwieldy as $r$ and $k$ increase.  For example, the case of $r=3$ and $k=10$ would require about $1.87 \times 10^7$ input magic states and a comparable number of quantum gates to distill $10^6$ output magic states.  However, since efficient multilevel distillation protocols, measured in the ratio of low-fidelity $\ket{H}$ input states consumed to yield a single high-fidelity $\ket{H}$ output, use $k \gg 1$ and multiple rounds, the greatest benefit from their application is seen in large-scale quantum computing, where a typical algorithm run may require $10^{12}$ magic states, each with error probability $10^{-12}$~\cite{Jones2012}.  Moreover, alternative designs can circumvent these issues.  If the first round uses a different protocol without correlated errors across logical magic states, such as Bravyi-Kitaev 15-to-1 distillation, then having multiple distillation blocks is unnecessary in the second round using a two-level concatenated protocol, which would lead to smaller multi-round, multilevel protocols.  Indeed, Sec.~\ref{analysis_section} shows that optimal protocols found by numerical search happen to take this approach.

The ``scaling exponent'' $\gamma$ of a distillation protocol characterizes its efficiency.  Specifically, $O(\log^{\gamma}(\epsilon_{\mathrm{in}}/\epsilon_{\mathrm{out}}))$ input states are required to distill one magic state of infidelity $\epsilon_{\mathrm{out}}$.  Scaling exponents for previous protocols are $\gamma \approx 2.46$ (``15-to-1''~\cite{Knill2004,Bravyi2005}), $\gamma \approx 2.32$ (``10-to-2''~\cite{Meier2012}), and $\gamma \approx 1.6$ (triorthogonal codes~\cite{Bravyi2012}).  Moreover, Bravyi and Haah conjecture that no magic-state distillation protocol has $\gamma < 1$~\cite{Bravyi2012}.  In this work, if each round of distillation uses one higher level of concatenation in the multilevel protocols, then the number of consumed inputs doubles.  In the limits of $k \rightarrow \infty$, $\epsilon \rightarrow 0$, multilevel protocols require $2^r + 1$ input states to each output state for $r$ rounds of distillation, where the $r^{\mathrm{th}}$ round is a level-$r$ distiller.  The final infidelity is $O((\epsilon_{\mathrm{in}})^{2^r})$, so the scaling exponent is $\gamma = \log(2^r + 1)/\log(2^r) \rightarrow 1$ as $r \rightarrow \infty$, which is the closest any protocol has come to reaching the conjectured bound.  We show later through numerical simulation that $\gamma \approx 1$ for error rates relevant to quantum computing.

\section{Analysis}
\label{analysis_section}
We make the conventional assumption that all quantum circuit operations are perfect, except for the initial $\ket{H}$ magic states we intend to distill.  This is a valid approximation if all operations are performed using fault-tolerant quantum error correction where the logical gate error is far below the final infidelity for distilled magic states~\cite{Preskill1998,Jones2012}; for a more explicit construction of fault-tolerant distillation circuits, see Ref.~\cite{Fowler2012}.  Additionally, following the methodology in Refs.~\cite{Bravyi2005,Meier2012}, we can consider each magic state with infidelity $\epsilon$ as the mixed state $\rho = (1-\epsilon)\ket{H}\bra{H} + \epsilon\ket{-H}\bra{-H}$, where $\ket{-H}$ is the $(-1)$ eigenstate of the Hadamard operation.

Determining the infidelity at the output of distillation becomes simply a matter of counting the distinct ways that errors lead to the circuit incorrectly accepting faulty states.  This process is aided by the geometric picture from earlier, and details are given in Appendix~\ref{analysis_appendix}.  It is essential that error probabilities $\epsilon_l$ and $\epsilon_p$ for each input magic state are independent.  Then a one-level, $(3k+8)$-to-$k$ distiller using the $H_{(k+4)}$ code has output error rate on each $\ket{H}$ state as
\begin{equation}
E_1^{(k+4)}(\epsilon_l,\epsilon_p) = (k-1){\epsilon_l}^2 + (2k+2){\epsilon_p}^2 + (\ldots),
\end{equation}
where higher order terms denoted (...) are omitted.  Our numerical results justify the use of lowest-order approximations as higher-order terms are negligible in optimally efficient protocols.  The lowest-order error rates are both second order, because the Hadamard basis measurement and $H_{(k+4)}$ code can together detect a single error in any magic state.  The probability of the distiller detecting an error, in which case the output is discarded, is $k\epsilon_l + 2(k+4)\epsilon_p + (\ldots)$.  If $\epsilon_l = \epsilon_p = \epsilon$, then the output error rate of $(3k+1)\epsilon^2$ conditioned on success is the same as in Ref.~\cite{Bravyi2012}.  Using the two-level distiller constructed from concatenated $H_{(k+4)}$ codes, the output infidelity for each distilled $\ket{H}$ state is
\begin{eqnarray}
& E_2^{(k+4)\times(k+4)}(\epsilon_l,\epsilon_p) = & (k^2 - 1){\epsilon_l}^2 + 8(k^2 + 4k + 3){\epsilon_p}^4 \nonumber \\
& & + (k+4)^2\epsilon_l {\epsilon_p}^2 + (\ldots).
\end{eqnarray}
The probability of the two-level distiller detecting an error is $k^2\epsilon_l + 2(k+4)^2\epsilon_p + 2k^2(k+4)^2 \epsilon_l\epsilon_p + (\ldots)$.  Similar error suppression extends to higher multilevel protocols, as examined in Appendix~\ref{analysis_appendix}.

\begin{figure}
  \centering
  \includegraphics[width=8.3cm]{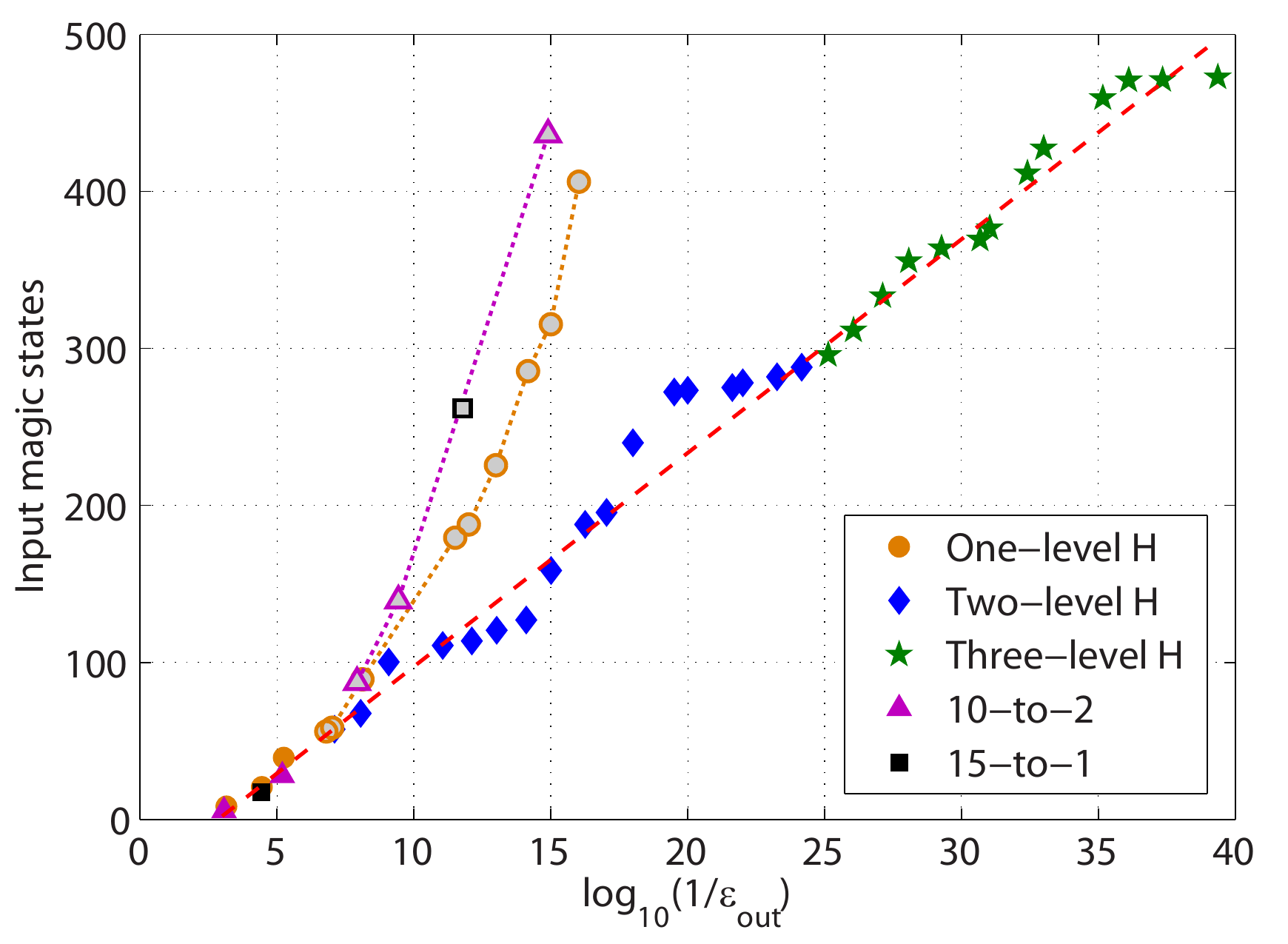}\\
  \caption{(Color online) Average number of input $\ket{H}$ states with $\epsilon_{\mathrm{in}} = 0.01$ consumed to produce a single output $\ket{H}$ state with fidelity $\epsilon_{\mathrm{out}}$.  Multiple-round distillation can use different protocols in each round, and the markers indicate just the last round of distillation.  The grey-shaded squares, triangles, and circles show, respectively, the best distillation possible with only \hbox{15-to-1}~\cite{Bravyi2005}, \hbox{10-to-2}~\cite{Meier2012}, and triorthogonal-code~\cite{Bravyi2012} protocols.  The dashed line is a linear fit $14\log_{10}(1/\epsilon_{\mathrm{out}}) - 40$.}
  \label{Protocols_plot}
\end{figure}

Figure~\ref{Protocols_plot} shows the performance of optimal multi-round distillation protocols identified by numerical search, indicating the number of input states with $\epsilon_0 = 0.01$ required to reach a desired output infidelity $\epsilon_{\mathrm{out}}$.  The markers indicate the type of protocol in the last round of distillation, including Bravyi-Kitaev~\cite{Bravyi2005}, Meier-Eastin-Knill~\cite{Meier2012}, and multilevel $H$~codes (see Appendix~\ref{numerical_appendix} for details).  The search attempts to identify the best distillation routines using any combination of known methods.  Note that the recent Bravyi-Haah protocols~\cite{Bravyi2012} have the same performance as one-level $H$~codes.  As expected, there is a trend of using higher-distance multilevel protocols in the last round as the output error rate $\epsilon_{\mathrm{out}}$ decreases (earlier rounds may use different protocols).  Where present, open markers indicate the best possible performance of previously studied protocols without the advent of multilevel distillation, and multilevel distillation is dominant for $\epsilon_{\mathrm{out}} \le 10^{-7}$, which is the regime pertinent to quantum computing.  Moreover, in this regime, input error rates are sufficiently small that only lowest-order terms in the $E(\cdot)$ output-error functions are significant.  The linear fit provides empirical evidence that the scaling exponent is $\gamma \approx 1$ in this regime, which demonstrates that multilevel protocols are close to the conjectured optimal performance in practice.

\section{Conclusions}
$H$~codes can distill magic states for $\texttt{T} = \exp(i \pi (I - \sigma^z)/8)$, which may enable distillation of three-qubit magic states for controlled-controlled-\texttt{Z}, which is locally equivalent to the Toffoli gate~\cite{Nielsen2000} (see Appendices~\ref{T_appendix} and~\ref{Toffoli_appendix} for details).  As a first pass at studying distillation protocols, this work considered only average input-to-output efficiency.  Future work will more rigorously examine to entire costs in qubits and gates required to fault-tolerantly distill magic states using multilevel codes~\cite{Fowler2013}. Multilevel distillation is an important development for large-scale, fault-tolerant quantum computing, where the distillation of magic states is often considered the most costly subroutine~\cite{Isailovic2008,Jones2012}.  Other codes with high density, high distance, and transversal Hadamard may yet be discovered, though for the present, $H$~codes are useful for their high efficiency and simple construction.

The author would like to thank Sergey Bravyi and Bryan Eastin for assistance in analyzing protocols at levels three and higher.  This work was supported by the Univ. of Tokyo Special Coordination Funds for Promoting Science and Technology, NICT, and the Japan Society for the Promotion of Science (JSPS) through its FIRST Program.

%\bibliography{References}

%

\appendix

\section{Error analysis in multilevel $H$-code distillation}
\label{analysis_appendix}
The multilevel codes analyzed here use concatenated $H$~codes.  When two $H$~codes are concatenated, the logical qubits of the first level of encoding are used as physical qubits for completely distinct codes at the second level.  Consider a two-level scheme: if the codes at first and second levels are $[[n_1,(n_1-4),2]]$ and $[[n_2,(n_2-4),2]]$, respectively, then the concatenated code is $[[n_1 n_2, (n_1-4)(n_2-4),4]]$, as shown in Fig.~2b of the main text. This process can be extended to higher levels of concatenation.

Determining the potential errors and their likelihood in multilevel protocols requires careful analysis.  Let us enumerate the error configurations which are detected by the protocols; the error probability is given by summing the probability of all error configurations that are not detected and that lead to error(s) in the encoded $\ket{H}$ states.  As a first step, we may simplify the analysis of multilevel codes by considering each input magic state to our quantum computer as having an independent probability of $\sigma^Y$ error, as discussed in Refs.~\cite{Bravyi2005,Meier2012}.  This allows us to consider only one type of error stemming from each magic state used in the protocol.

Identifying undetected error events in multilevel distillation, which lead to output error rate, is aided by the geometric picture introduced in the main text.  Qubits which will form the code are arranged in a rectangular solid, then grouped in lines along each dimension for encoding.  There are two error-detecting steps which together implement distillation: the Hadamard-basis measurement and the error detection of the $H$~codes.  The Hadamard measurement registers an error for odd parity in the total of encoded state errors and physical-level errors in the first round of $R_Y(-\pi/4)$ gates, and there is one of these for each qubit site in the code (see Fig.~1 of the main text).

The second method for $H$~codes to detect errors is by measuring the code stabilizers.  The code stabilizers detect any configuration of errors which is not a logical operator in the concatenated code.  Because of the redundant structure using overlapping $H$~codes, only a very small fraction of error configurations evade detection.  Before moving on, note that at each qubit site, there are two faulty gates applied, and two errors on the same qubit will cancel (however, the first error will propagate to the Hadamard-basis measurement).  Conversely, a single error in one of the two gates will propagate to the stabilizer-measurement round, but only an error in the first gate will also propagate to the Hadamard measurement.  The stabilizer-measurement round will only ``see'' the odd/even parity of the number of errors at each qubit site.

One type of error event that occurs at concatenation levels three and higher requires special treatment.  If there is an error in an encoded magic state and errors on two physical states used for the same controlled-Hadamard gate at the physical level, then this combination of input errors is not detected by the distillation protocol, leading to logical output error.  This event leads to the $O(\epsilon_l{\epsilon_p}^2)$ error probability from the main text, which is not an issue for two-level protocols, but it must be addressed in levels three and higher.  The solution for $t$-level distillation, where $t \ge 3$, is to repeat the controlled-Hadamard measurement $2^{(t-2)}$ times, consuming $2^{(t-1)}$ magic states at the physical level.  After each transversal controlled-Hadamard, the code syndrome checks for detectable error patterns.  With this procedure, one encoded-state error would also require at least $2^{(t-1)}$ errors in physical-level magic states to go undetected, leading to probability of error that scales as $O(\epsilon_l{\epsilon_p}^{2^{t-1}})$.

Consider the pattern of errors after the two potentially faulty gates on each qubit in the $t$-dimensional cartesian grid arrangement.  The many levels of error checking in the $H$ codes can detect a single error in any encoding block at any encoding level.  For this analysis, let us separate the $(k+4)$ qubits in a single $H$~code block into two groups: the first four qubits are ``preamble'' qubits, while the remaining $k$ qubits are index qubits.  The reason for distinction is that the logical $\overline{\texttt{Y}}_i$ operators, which would also be undetected error configurations, have common physical-qubit operators in the preamble, with a degeneracy of two: $\overline{\texttt{Y}}_i = -\texttt{Y}_1 \texttt{Y}_3 \texttt{Y}_{i+4} = -\texttt{Y}_2 \texttt{Y}_4 \texttt{Y}_{i+4}$, because of the stabilizer $\texttt{Y}_1 \texttt{Y}_2 \texttt{Y}_3 \texttt{Y}_4$.  Conversely, the logical operators are distinguished by the $i^{\mathrm{th}}$ logical Pauli operator having a physical Pauli operator on the $i^{\mathrm{th}}$ index qubit (numbered $(i+4)$ when preamble is included).

With the preamble/index distinction, we can now identify the most likely error patterns.  For any size $H$~code, there are two weight-2 errors in the preamble: $\texttt{Y}_1 \texttt{Y}_2$ and $\texttt{Y}_3 \texttt{Y}_4$.  Logically, these represent the product of $\overline{\texttt{Y}}$ operators on all encoded qubits.  In the index qubits, any pair of errors is logical: $\texttt{Y}_{i+4} \texttt{Y}_{j+4} = \overline{\texttt{Y}}_{i} \overline{\texttt{Y}}_j$.  However, a pair of errors split with one each in preamble and index is always detectable by the code stabilizers.  Thus, any single encoded qubit could have a logical error stemming from a pair of errors in two different configurations in the preamble or $(k-1)$ configurations in the index qubits.  There is also one weight-three error.  Each physical-state error configuration is multiplied by a degeneracy factor that is the number of ways an even number of errors occur before the \texttt{CNOT} in Fig.~1, thereby evading the Hadamard measurement.  Thus the probability of logical error is $2(k+1){\epsilon_p}^2 + 4{\epsilon_p}^3 + O({\epsilon_p}^4)$.  The Hadamard measurement fails to detect an even number of errors in the logical input states.  There are $(k-1)$ ways that a pair of encoded input errors could corrupt any given qubit and $(k-1)(k-2)(k-3)/6$ ways four errors could corrupt any given qubit (assuming $k \ge 4$).  This contributes error terms $(k-1){\epsilon_l}^2 + (1/6)(k-1)(k-2)(k-3){\epsilon_l}^4 + O({\epsilon}^6)$. Finally, it is possible for a single logical error and an odd number of physical errors before the \texttt{CNOT} in Fig.~1 of the main text, potentially in conjunction with other physical errors after \texttt{CNOT}, to occur simultaneously in a way that evades both checks.  This contributes a term $(k+4)\epsilon_l{\epsilon_p}^2 + 8(k-1)\epsilon_l{\epsilon_p}^3 + O(\epsilon_l{\epsilon_p}^4)$.

The numerical analysis detailed below shows that efficient use of one-level $H$~codes has similar error rates for $\epsilon_l$ and $\epsilon_p$, and both are below 0.01, so the relevant terms in the error functions for one-level $H$~codes are $E_1^{(k+4)}(\epsilon_l,\epsilon_p) = (k-1){\epsilon_l}^2 + (2k+2){\epsilon_p}^2 + (\ldots)$, which reproduces Eqn.~(1) of the main text.  As a result, the higher-order terms above can be neglected for this range of parameters so long as $k$ is not too large.  Simply put, if the higher terms become relevant (\emph{i.e.} $\epsilon_l$, $\epsilon_p$, or $k$ is sufficiently large in magnitude), then the distillation protocol is being used ineffectively, and it may in fact cause more errors than it corrects.  These findings are supported by the numerical search for optimal protocols, and we proceed using this approximation.

When $H$~codes are concatenated, the analysis of undetected error patterns becomes more complicated.  In particular, logical errors from one layer of encoding must be ``matched'' with errors from other encoding blocks to go undetected at the next level.  Consider the case of the level-two-concatenated, square-array distiller, and focus on one of the encoded states.  As before, a pair of encoded-state input errors evades the Hadamard measurement, which contributes a term $(k^2 - 1){\epsilon_l}^2$.  The undetected errors resulting from consumed magic states are more complicated.  Within the upper encoding block, there are two ways a logical error could be caused by a pair of errors in the preamble, and $(k-1)$ possibilities for logical error from a pair of index errors.  However, each of the inputs to the second level are the logical qubits of distinct $H$~codes at the first level, which has additional error detection.  The most likely errors from the first level come in pairs, but these pairs are sent to different codes at the second level.  As a result, the error patterns from the first level must come in ``matched'' pairs that are also not detected at the second level.  For any particular error configuration going into a block at the second level, there are four preamble configurations and $(k-1)$ index configurations at the first level that could have caused it.  There are $(k+1)$ undetected error configurations at the second level, and the degeneracy factor of four physical errors is eight, so the consumed magic states contribute a term $8(k+1)(k+3){\epsilon_p}^4$.  Finally, the most likely way that physical and encoded errors can occur in conjunction is a logical error on the magic state in question and two physical errors on the same qubit anywhere, which has probability $(k+4)^2 \epsilon_l {\epsilon_p}^2$.  Combined, these error terms reproduce the results in Eqn.~(2): $E_2^{(k+4)\times(k+4)}(\epsilon_l,\epsilon_p) = (k^2 - 1){\epsilon_l}^2 + 8(k^2 + 4k + 3){\epsilon_p}^4 + (k+4)^2\epsilon_l {\epsilon_p}^2 + (\ldots)$.  We drop terms at higher order because they are found to be negligible in optimal protocols.  For example, the first optimal two-level protocol has parameters $k = 8$, $\epsilon_l = 3.5 \times 10^{-5}$, and $\epsilon_p = 9 \times 10^{-4}$, where both input types come from earlier rounds of distillation (Bravyi-Kitaev and Meier-Eastin-Knill, respectively).  More details of the numerical search are given below.

Continuing this approach, one can show the significant error terms at level $t \ge 3$ are given by
\begin{eqnarray}
& E_t^{(k+4)^t}&(\epsilon_l,\epsilon_p) = (k^t - 1){\epsilon_l}^2 \nonumber \\
& & + 2^{(2^t + t-3)}(k+1)(k+3)^{t-1} {\epsilon_p}^{(2^t)} \nonumber \\
& & + (k+4)^{t(2^{(t-2)})}\epsilon_l {\epsilon_p}^{(2^{(t-1)})} \nonumber \\
& & + (\ldots).
\end{eqnarray}
These terms incorporate degeneracy in error configurations and repeated Hadamard measurements.  The coefficients of the second and third terms on the RHS of Eqn.~(1) represent  physical error configurations and encoded/physical combinations, respectively, and these grow rapidly as a function of $r$.  Accordingly, the optimal-protocol search does not advocate the use of three-level protocols until the desired output error rate is below $10^{-25}$, which is beyond the needs of any quantum algorithm so far conceived.  No four-level protocols were found to be optimal for output error rates above $10^{-40}$, which under practical considerations means they are not likely to ever be used.  The next section considers the size of multilevel distillation circuits, which can also limit their usefulness.

\section{Optimal multi-round distillation}
\label{numerical_appendix}
The claimed efficiency of multilevel distillation was examined quantitatively with a numerical search for optimal multi-round distillation protocols.  Each of the protocols is optimal in the sense that, for a given final infidelity $\epsilon_{\mathrm{out}}$, no other sequence requires fewer average input states, and, for a given average number of input states, no other protocol achieves lower $\epsilon_{\mathrm{out}}$.  Note that probability of rejection upon detected error is incorporated by considering average cost for distillation when failure-and-repeat steps are included.  The protocols plotted in Fig.~\ref{Protocols_plot} of the main text are just the last round of a distillation sequence.  Earlier rounds can be, and usually are, different protocols.  The search space was constrained such that the number of rounds $r \le 5$, number of encoded logical qubits $k \le 20$ for all $H$~codes, and multilevel codes are square $(k+4) \times (k+4)$, \emph{etc.}

Generally speaking, smaller protocols handle large input error rates in early rounds better, while larger multilevel protocols are more inefficient at distillation when input error rates are low enough.  For example, the protocols listed in Table~\ref{optimal_protocols_table} are the same ones plotted in Fig.~~\ref{Protocols_plot}.  Note that when $\epsilon_{\mathrm{out}}$ is smaller than $10^{-7}$, multilevel protocols are most efficient.  Should one desire $\epsilon_{\mathrm{out}} < 10^{-25}$, level-three protocols have the highest efficiency.  Higher levels of concatenation (up to level five) were part of the search, but they were not efficient for $\epsilon_{\mathrm{out}} > 10^{-40}$.  The error-function notation $E(\cdot)$ from the main text is used to show how the inputs to later rounds of distillation may be the outputs of an earlier round.  This notation neglects the total size of the distillers, which must be determined by using parallel distillation blocks whenever correlated errors on logical magic states are present.

For reference, the best achievable results with prior protocols are also shown in Table~\ref{optimal_protocols_table}, in reverse chronological order of their discovery.  The methods are cumulative, so the most recent Bravyi-Haah codes (2012) could also use Meier-Eastin-Knill (2012) or Bravyi-Kitaev (2005) distillation, but the oldest Bravyi-Kitaev distillation is alone in its column.  The best achievable results with older protocols are also shown in Fig.~\ref{Protocols_plot} for comparison.  Accordingly, the multilevel protocols can use any of the above protocols wherever the numerical search finds doing so to be optimally efficient.

The numerical simulation uses error functions $E(\cdot)$ for the Bravyi-Kitaev (``BK,''~\cite{Bravyi2005}) \hbox{15-to-1}  and Meier-Eastin-Knill (``MEK,''~\cite{Meier2012}) 10-to-2 protocols.  The first three Taylor-series terms of these functions near $\epsilon = 0$ are:
\begin{equation}
E_{\mathrm{BK}}(\epsilon) = 35\epsilon^3 + 105\epsilon^4 + 378\epsilon^5 + (\ldots);
\end{equation}
\begin{equation}
E_{\mathrm{MEK}}(\epsilon) = 9\epsilon^2 - 56\epsilon^3 + 160\epsilon^4 + (\ldots).
\end{equation}
In the numerical search, $\epsilon \le 0.01$, so the first term for each dominates.  The Bravyi-Haah triorthogonal codes (``BH,''~\cite{Bravyi2012}) have the same error function as $H$~codes (\emph{cf}. Eqn.~(1) of the main text):
\begin{equation}
E_{\mathrm{BH}}^{(k)}(\epsilon) = E_1^{(k+4)}(\epsilon,\epsilon) = (3k+1){\epsilon}^2 + (\ldots).
\end{equation}
This correspondence, combined with the results of Reichardt~\cite{Reichardt2005}, suggests a connection between these code families.

\section{Distilling magic states for $T$ gates with $H$~codes}
\label{T_appendix}
In addition to distilling Hadamard states $\ket{H}$, $H$~codes can also distill the magic state \hbox{$\ket{A} = (1/\sqrt{2})(\ket{0} + e^{i\pi/4}\ket{1})$}, which is used to make the $\texttt{T} = \exp(i\pi(I-\sigma^z)/8)$ gate.  This construction is useful by itself, but it can also be used to make Toffoli magic states as shown in the next section.  State $\ket{A}$ is stabilized by the operator $\texttt{T}\texttt{X}\texttt{T}^{\dag} = (1/\sqrt{2})(\texttt{X} + \texttt{Y})$, which is also transversal in $H$ codes.  Distillation is performed by encoding $\ket{A}$ states as logical qubits, then measuring the controlled-$(\texttt{T}\texttt{X}\texttt{T}^{\dag})$ using $\ket{A}$ states at the physical level, followed by routine error detection.

\begin{figure}
  \includegraphics[width=8.3cm]{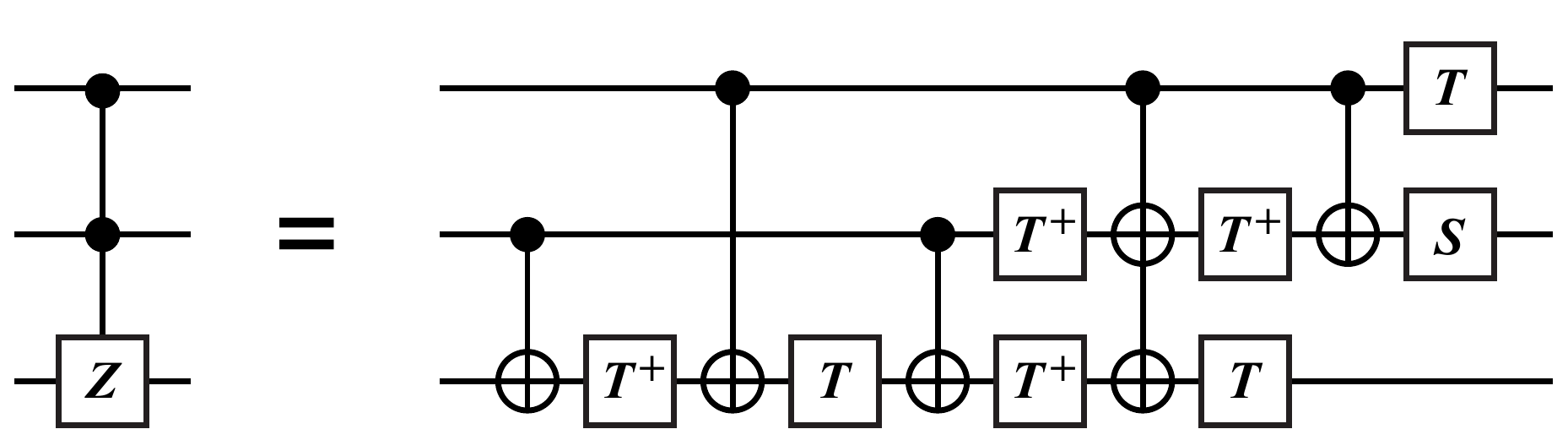}\\
  \caption{A controlled-controlled-\texttt{Z} gate, which is locally equivalent to a Toffoli, can be decomposed into \texttt{CNOT} and \texttt{T}~gates.  A CSS+T code has transversal \texttt{CNOT} and \texttt{T}, so it could be used to distill three-qubit magic states for the Toffoli gate.}
  \label{CCZ_circuit}
\end{figure}

\section{Distilling Toffoli magic states}
\label{Toffoli_appendix}
A CSS quantum code~\cite{Calderbank1996,Steane1996} necessarily has a transversal \texttt{CNOT} operation.  A CSS code with transversal \texttt{T}~operation (let us use the shorthand ``CSS+T'') will also have a transversal controlled-controlled-\texttt{Z} (CCZ) operation, because the latter quantum gate can be decomposed into \texttt{CNOT} and~\texttt{T} (or $\texttt{T}^{\dag}$), as shown in Fig.~\ref{CCZ_circuit} of this supplementary information.  The CCZ gate is locally equivalent to Toffoli (via Hadamard transforms on the target qubit), so CSS+T codes can distill a magic state for the CCZ gate, which is equivalent to distilling Toffoli magic states (see p.~488 of Ref.~\cite{Nielsen2000}), assuming Clifford operations are freely available.

\cleardoublepage

\begin{longtable*}{|>{\centering\arraybackslash}m{2.2cm}|>{\centering\arraybackslash}m{2cm}|m{8cm}|>{\centering\arraybackslash}m{1cm}|>{\centering\arraybackslash}m{1cm}|>{\centering\arraybackslash}m{1cm}|>{\centering\arraybackslash}m{1cm}|}
  %\centering
  %\begin{tabular}{|>{\centering\arraybackslash}m{2.2cm}|>{\centering\arraybackslash}m{2cm}|m{8cm}|>{\centering\arraybackslash}m{1cm}|>{\centering\arraybackslash}m{1cm}|>{\centering\arraybackslash}m{1cm}|>{\centering\arraybackslash}m{1cm}|}
    \hline
    $-\log_{10}(\epsilon_{\mathrm{target}})$ & $-\log_{10}(\epsilon_{\mathrm{out}})$ & Protocol & $C_{\mathrm{ML}}$ & $C_{\mathrm{BH}}$ & $C_{\mathrm{MEK}}$ & $C_{\mathrm{BK}}$ \\ \hline
    4 & 4.46 & $E_{\mathrm{BK}}(\epsilon_0)$ & 17.44 & 17.44 & 17.44 & 17.44 \\
    5 & 5.14 & $E_{\mathrm{MEK}}(E_{\mathrm{BK}}(\epsilon_0))$ & 27.93 & 27.86 & 27.86 & 261.5 \\
    6 & 6.83 & $E_{\mathrm{BH}}^{(40)}(E_{\mathrm{BK}}(\epsilon_0))$ & 56.07 & 56.07 & 83.99 & 261.5 \\
    7 & 7.11 & $E_2^{(12\times12)}(E_{\mathrm{BK}}(\epsilon_0),E_{\mathrm{MEK}}(\epsilon_0))$  & 57.38 & 58.30 & 83.99 & 261.5 \\
    8 & 8.06 & $E_2^{(10\times10)}(E_{\mathrm{MEK}}(E_{\mathrm{MEK}}(\epsilon_0)),E_{\mathrm{MEK}}(\epsilon_0))$ & 67.52 & 89.26 & 139.3 & 261.5 \\
    9 & 9.08 & $E_2^{(10\times10)}(E_{\mathrm{MEK}}(E_{\mathrm{BH}}^{(2)}(\epsilon_0)),E_{\mathrm{BH}}^{(2)}(\epsilon_0))$ & 100.3 & 139.3 & 139.3 & 261.5 \\
    10 & 11.1 & $E_2^{(24\times24)}(E_{\mathrm{BH}}^{(40)}(E_{\mathrm{BK}}(\epsilon_0)),E_{\mathrm{BK}}(\epsilon_0))$ & 110.7 & 179.4 & 261.7 & 261.5 \\
    11 & 11.1 & ---same as above--- & 110.7 & 179.4 & 261.7 & 261.5 \\
    12 & 12.1 & $E_2^{(24\times24)}(E_2^{(10\times10)}(E_{\mathrm{BK}}(\epsilon_0),E_{\mathrm{MEK}}(\epsilon_0)),$ \newline $E_{\mathrm{BK}}(\epsilon_0))$ & 113.7 & 187.9 & 418.0 & 3923. \\
    13 & 13.0 & $E_2^{(24\times24)}(E_2^{(10\times10)}(E_{\mathrm{BH}}^{(8)}(E_{\mathrm{MEK}}(\epsilon_0)))),$ \newline $E_{\mathrm{MEK}}(\epsilon_0)),E_{\mathrm{BK}}(\epsilon_0))$ & 120.4 & 225.6 & 418.0 & 3923. \\
    14 & 14.1 & $E_2^{(24\times24)}(E_2^{(10\times10)}(E_{\mathrm{MEK}}(E_{\mathrm{MEK}}(\epsilon_0)))),$ \newline $E_{\mathrm{MEK}}(\epsilon_0)),E_{\mathrm{BK}}(\epsilon_0))$ & 126.9 & 285.6 & 419.9 & 3923. \\
    15 & 15.0 & $E_2^{(14\times14)}(E_2^{(10\times10)}(E_{\mathrm{MEK}}(E_{\mathrm{MEK}}(\epsilon_0)))),$ \newline $E_{\mathrm{MEK}}(\epsilon_0)),E_{\mathrm{BH}}^{(10)}(E_{\mathrm{MEK}}(\epsilon_0)))$ & 158.5 & 315.5 & 696.7 & 3923. \\
    16 & 16.3 & $E_2^{(24\times24)}(E_2^{(10\times10)}(E_{\mathrm{BH}}^{(40)}(E_{\mathrm{BK}}(\epsilon_0)),$ \newline $E_{\mathrm{MEK}}(\epsilon_0)),E_{\mathrm{MEK}}(E_{\mathrm{MEK}}(\epsilon_0)))$ & 187.9 & 406.2 & 696.7 & 3923. \\
    17 & 17.0 & $E_2^{(22\times22)}(E_2^{(24\times24)}(E_{\mathrm{BH}}^{(40)}(E_{\mathrm{BK}}(\epsilon_0)),$ \newline $E_{\mathrm{BK}}(\epsilon_0)),E_{\mathrm{MEK}}(E_{\mathrm{MEK}}(\epsilon_0)))$ & 195.5 & 529.5 & 696.7 & 3923. \\
    18 & 18.0 & $E_2^{(20\times20)}(E_2^{(24\times24)}(E_{\mathrm{BH}}^{(38)}(E_{\mathrm{BK}}(\epsilon_0)),$ \newline $E_{\mathrm{BK}}(\epsilon_0)),E_{\mathrm{MEK}}(E_{\mathrm{BH}}^{(40)}(\epsilon_0)))$ & 239.8 & 574.1 & 1260. & 3923. \\
    19 & 19.5 & $E_2^{(24\times24)}(E_2^{(24\times24)}(E_{\mathrm{BH}}^{(40)}(E_{\mathrm{BK}}(\epsilon_0)),$ \newline $E_{\mathrm{BK}}(\epsilon_0)),E_{\mathrm{BH}}^{(40)}(E_{\mathrm{BK}}(\epsilon_0)))$ & 272.1 & 574.1 & 1260. & 3923. \\
    20 & 20.0 & $E_2^{(24\times24)}(E_2^{(24\times24)}(E_{\mathrm{BH}}^{(30)}(E_{\mathrm{BK}}(\epsilon_0)),$ \newline $E_{\mathrm{BK}}(\epsilon_0)),E_{\mathrm{BH}}^{(40)}(E_{\mathrm{BK}}(\epsilon_0)))$ & 273.3 & 574.1 & 1260. & 3923. \\
    21 & 21.6 & $E_2^{(24\times24)}(E_2^{(24\times24)}(E_2^{(10\times10)}(E_{\mathrm{BK}}(\epsilon_0),$ \newline $E_{\mathrm{MEK}}(\epsilon_0)),E_{\mathrm{BK}}(\epsilon_0)),E_{\mathrm{BH}}^{(40)}(E_{\mathrm{BK}}(\epsilon_0)))$ & 275.1 & 575.9 & 1260. & 3923. \\
    22 & 22.0 & $E_2^{(24\times24)}(E_2^{(20\times20)}(E_2^{(10\times10)}(E_{\mathrm{BK}}(\epsilon_0),$ \newline $E_{\mathrm{MEK}}(\epsilon_0)),E_{\mathrm{BK}}(\epsilon_0)),E_{\mathrm{BH}}^{(40)}(E_{\mathrm{BK}}(\epsilon_0)))$ & 278.0 & 604.3 & 1308. & 3923. \\
    23 & 23.3 & $E_2^{(24\times24)}(E_2^{(24\times24)}(E_2^{(10\times10)}(E_{\mathrm{BH}}^{(8)}(E_{\mathrm{MEK}}(\epsilon_0)),$ \newline $E_{\mathrm{MEK}}(\epsilon_0)),E_{\mathrm{BK}}(\epsilon_0)),E_{\mathrm{BH}}^{(40)}(E_{\mathrm{BK}}(\epsilon_0)))$ & 281.9 & 652.3 & 2090. & 3923. \\
    24 & 24.2 & $E_2^{(24\times24)}(E_2^{(24\times24)}(E_2^{(10\times10)}(E_{\mathrm{BH}}^{(6)}(E_{\mathrm{MEK}}(\epsilon_0)),$ \newline $E_{\mathrm{MEK}}(\epsilon_0)),E_{\mathrm{BK}}(\epsilon_0)),E_2^{(12\times12)}(E_{\mathrm{BK}}(\epsilon_0),E_{\mathrm{MEK}}(\epsilon_0)))$ & 287.9 & 731.5 & 2090. & 3923. \\
    25 & 25.1 & $E_3^{(16\times16\times16)}(E_2^{(22\times22)}(E_2^{(10\times10)}(E_{\mathrm{MEK}}(E_{\mathrm{MEK}}(\epsilon_0)),$ \newline $E_{\mathrm{MEK}}(\epsilon_0)),E_{\mathrm{BK}}(\epsilon_0)),E_{\mathrm{MEK}}(E_{\mathrm{MEK}}(\epsilon_0)))$ & 295.7 & 853.1 & 2090. & 3923. \\
    26 & 26.1 & $E_3^{(16\times16\times16)}(E_2^{(14\times14)}(E_2^{(10\times10)}(E_{\mathrm{MEK}}(E_{\mathrm{MEK}}(\epsilon_0)),$ \newline $E_{\mathrm{MEK}}(\epsilon_0)),E_{\mathrm{BK}}(\epsilon_0)),E_{\mathrm{MEK}}(E_{\mathrm{MEK}}(\epsilon_0)))$ & 311.5 & 914.0 & 2090. & 3923. \\
    27 & 27.1 & $E_3^{(16\times16\times16)}(E_2^{(24\times24)}(E_2^{(10\times10)}(E_{\mathrm{MEK}}(E_{\mathrm{BH}}^{(2)}(\epsilon_0)),$ \newline $E_{\mathrm{MEK}}(\epsilon_0)),E_{\mathrm{BH}}^{(6)}(E_{\mathrm{MEK}}(\epsilon_0))),E_{\mathrm{MEK}}(E_{\mathrm{MEK}}(\epsilon_0)))$ & 333.3 & 947.5 & 2100. & 3923. \\
    28 & 28.1 & $E_3^{(16\times16\times16)}(E_2^{(18\times18)}(E_2^{(10\times10)}(E_{\mathrm{MEK}}(E_{\mathrm{BH}}^{(2)}(\epsilon_0)),$ \newline $E_{\mathrm{MEK}}(\epsilon_0)),E_{\mathrm{MEK}}(E_{\mathrm{MEK}}(\epsilon_0))),E_{\mathrm{MEK}}(E_{\mathrm{MEK}}(\epsilon_0)))$ & 355.6 & 1015. & 2181. & 3923. \\
    29 & 29.3 & $E_3^{(16\times16\times16)}(E_2^{(18\times18)}(E_2^{(10\times10)}(E_{\mathrm{BH}}^{(40)}(E_{\mathrm{BK}}(\epsilon_0)),$ \newline $E_{\mathrm{MEK}}(\epsilon_0)),E_{\mathrm{MEK}}(E_{\mathrm{MEK}}(\epsilon_0))),E_{\mathrm{MEK}}(E_{\mathrm{MEK}}(\epsilon_0)))$ & 363.7 & 1125. & 3483. & 3923. \\
    30 & 30.7 & $E_3^{(16\times16\times16)}(E_2^{(24\times24)}(E_2^{(24\times24)}(E_{\mathrm{BH}}^{(40)}(E_{\mathrm{BK}}(\epsilon_0)),$ \newline $E_{\mathrm{BK}}(\epsilon_0)),E_{\mathrm{MEK}}(E_{\mathrm{MEK}}(\epsilon_0))),E_{\mathrm{MEK}}(E_{\mathrm{MEK}}(\epsilon_0)))$ & 369.3 & 1301. & 3483. & 3923. \\
    31 & 31.0 & $E_3^{(16\times16\times16)}(E_2^{(20\times20)}(E_2^{(24\times24)}(E_{\mathrm{BH}}^{(40)}(E_{\mathrm{BK}}(\epsilon_0)),$ \newline $E_{\mathrm{BK}}(\epsilon_0)),E_{\mathrm{MEK}}(E_{\mathrm{MEK}}(\epsilon_0))),E_{\mathrm{MEK}}(E_{\mathrm{MEK}}(\epsilon_0)))$ & 376.5 & &   & 3923. \\
    32 & 32.4 & $E_3^{(16\times16\times16)}(E_2^{(24\times24)}(E_2^{(24\times24)}(E_{\mathrm{BH}}^{(40)}(E_{\mathrm{BK}}(\epsilon_0)),$ \newline $E_{\mathrm{BK}}(\epsilon_0)),E_{\mathrm{MEK}}(E_{\mathrm{BH}}^{(2)}(\epsilon_0))),E_{\mathrm{MEK}}(E_{\mathrm{MEK}}(\epsilon_0)))$ & 411.5 & &   & 3923. \\
    33 & 33.0 & $E_3^{(14\times14\times14)}(E_2^{(20\times20)}(E_2^{(24\times24)}(E_{\mathrm{BH}}^{(38)}(E_{\mathrm{BK}}(\epsilon_0)),$ \newline $E_{\mathrm{BK}}(\epsilon_0)),E_{\mathrm{MEK}}(E_{\mathrm{BH}}^{(2)}(\epsilon_0))),E_{\mathrm{MEK}}(E_{\mathrm{MEK}}(\epsilon_0)))$ & 427.3 & &   & 3923. \\
    34 & 35.2 & $E_3^{(16\times16\times16)}(E_2^{(24\times24)}(E_2^{(24\times24)}(E_{\mathrm{BH}}^{(40)}(E_{\mathrm{BK}}(\epsilon_0)),$ \newline $E_{\mathrm{BK}}(\epsilon_0)),E_{\mathrm{BH}}^{(40)}(E_{\mathrm{BK}}(\epsilon_0))),E_{\mathrm{MEK}}(E_{\mathrm{MEK}}(\epsilon_0)))$ & 459.4 & &   & 58838. \\
    35 & 35.2 & ---same as above--- & 459.4 & &   & 58838. \\
    36 & 36.1 & $E_3^{(24\times24\times24)}(E_2^{(24\times24)}(E_2^{(24\times24)}(E_{\mathrm{BH}}^{(30)}(E_{\mathrm{BK}}(\epsilon_0)),$ \newline $E_{\mathrm{BK}}(\epsilon_0)),E_{\mathrm{BH}}^{(40)}(E_{\mathrm{BK}}(\epsilon_0))),E_{\mathrm{BH}}^{(40)}(E_{\mathrm{BK}}(\epsilon_0)))$ & 470.8 & &   & 58838. \\
    37 & 37.3 & $E_3^{(24\times24\times24)}(E_2^{(24\times24)}(E_2^{(24\times24)}(E_2^{(12\times12)}(E_{\mathrm{BK}}(\epsilon_0),$ \newline $E_{\mathrm{MEK}}(\epsilon_0)),E_{\mathrm{BK}}(\epsilon_0)),E_{\mathrm{BH}}^{(40)}(E_{\mathrm{BK}}(\epsilon_0))),E_{\mathrm{BH}}^{(40)}(E_{\mathrm{BK}}(\epsilon_0)))$ & 471.0 & &   & 58838. \\
    38 & 39.4 & $E_3^{(24\times24\times24)}(E_2^{(24\times24)}(E_2^{(24\times24)}(E_2^{(10\times10)}(E_{\mathrm{BK}}(\epsilon_0),$ \newline $E_{\mathrm{MEK}}(\epsilon_0)),E_{\mathrm{BK}}(\epsilon_0)),E_{\mathrm{BH}}^{(40)}(E_{\mathrm{BK}}(\epsilon_0))),E_{\mathrm{BH}}^{(40)}(E_{\mathrm{BK}}(\epsilon_0)))$ & 472.6 & &   & 58838. \\
    39 & 39.4 & ---same as above--- & 472.6 & &   & 58838. \\
    \hline
  %\end{tabular}
  \caption{Optimal distillation protocols identified by numerical search.  Protocols are specified using the error functions $E(\cdot)$ to indicate when inputs to one round are the outputs of another distillation circuit.  The data for $C_{\mathrm{BH}}$ and $C_{\mathrm{MEK}}$ are obtained from Ref.~\cite{Bravyi2012}.}
  \label{optimal_protocols_table}
\end{longtable*}

\end{document}